\documentclass[reprint,showpacs,preprintnumbers,pre,superscriptaddress]{revtex4-1}
\usepackage{graphicx}
\begin{document}

\title{Weighted reciprocal of temperature, weighted thermal flux, and their applications in finite-time thermodynamics}
\author{Shiqi Sheng}
\affiliation{Department of Physics, Beijing Normal University, Beijing 100875, China}
\author{Z. C. Tu}\email[Corresponding author. Email: ]{tuzc@bnu.edu.cn}
\affiliation{Department of Physics, Beijing Normal University, Beijing 100875, China}
\affiliation{Kavli Institute for Theoretical Physics China, CAS, Beijing 100190, China}


\begin{abstract}The concepts of weighted reciprocal of temperature and weighted thermal flux
are proposed for a heat engine operating between two heat baths and outputting mechanical work. With the aid of these two concepts, the generalized thermodynamic fluxes and forces can be expressed in a consistent way within the framework of irreversible thermodynamics. Then the efficiency at maximum power output for a heat engine, one of key topics in finite-time thermodynamics, is investigated on the basis of a generic model under the tight-coupling condition. The corresponding results have the same forms as those of low-dissipation heat engines [M. Esposito, R. Kawai, K. Lindenberg, and C. Van den Broeck, {Phys. Rev. Lett.} \textbf{105}, 150603 (2010)]. The mappings from two kinds of typical heat engines, such as the low-dissipation heat engine and the Feynman ratchet, into the present generic model are constructed. The universal efficiency at maximum power output up to the quadratic order is found to be valid for a heat engine coupled symmetrically and tightly with two baths. The concepts of weighted reciprocal of temperature and weighted thermal flux are also transplanted to the optimization of refrigerators.
\end{abstract}
\preprint{Phys. Rev. E \textbf{89}, 012129 (2014)}
\maketitle

\section{Introduction}
Entropy production rate is a crucial physical quantity in irreversible thermodynamics~\cite{Prigoginebook}. This quantity can be expressed in a canonical form that is the sum of products of generalized thermodynamic fluxes and forces.
When a system is in contact with two heat baths, the heat absorbed from one bath per unit time is equal to that released into another bath per unit time if no work exchange takes place between the system and the environment. Thus the thermodynamic flux related to the heat transfer between two baths, which is briefly called the thermal flux, can be taken as either the absorbed heat or released heat per unit time within the framework of irreversible thermodynamics~\cite{Prigoginebook}. However, the absorbed heat is unequal to the released heat if the system (for example, a heat engine or a refrigerator) outputs or inputs mechanical work. Therefore, we face a difficult circumstance when we construct the thermal flux for the heat engine or the refrigerator. There exists some arbitrariness when we choose either the absorbed heat or released heat per unit time as the thermal flux. To eliminate this arbitrariness, Jarzynski and Mazonka~\cite{Jarzynski99} intuitively suggested adopting the mean value of the absorbed heat and released heat per unit time as the thermal flux when they discussed a simple, discrete model of the Feynman ratchet operating between two heat baths. This convention was also followed by Nakagawa and Komatsu~\cite{Nakagawa2006} when they discussed the Feynman's ratchet as a heat pump. However, there is still a lack of a rigorous and systematic scheme to eliminate this arbitrariness.

The other classic topic on nonequilibrium processes is finite-time thermodynamics~\cite{Chambadal,Novikov,Curzon1975,devos85,Chen1989,ChenJC94,Bejan96,YanChen1990,RocoPRE12,Lingenchen1995,Velasco1997,Jincan1998,dcisbj2007,vdbrk2005,Berry1977,Berry1982,Berry1985,Gordon1991} which is concerned mainly with the energy conversion efficiency for heat devices including heat engines and refrigerators that complete thermodynamic cycles in finite time or operate in finite net rate. The most celebrated result in the finite-time thermodynamics is the efficiency at maximum power output for endoreversible heat engines~\cite{Novikov,Chambadal,Curzon1975}, $\eta_{CA}\equiv 1-\sqrt{1-\eta_C}$ with $\eta_C$ representing the Carnot efficiency. Interestingly, the corresponding result was also obtained by Yan and Chen for endoreversible refrigerators~\cite{YanChen1990}. They suggested taking the product of the coefficient of performance (COP) and the heat absorbed by the working substance from the cold bath per unit time as the optimization target function, which was also called the $\chi-$criterion by de Tom\'{a}s \emph{et al.} in recent work~\cite{RocoPRE12}. The COP at maximum $\chi-$criterion was found to be $\varepsilon_{YC}\equiv\sqrt{\varepsilon_C+1}-1$ for endoreversible refrigerators~\cite{YanChen1990} or symmetrically low-dissipation refrigerators~\cite{RocoPRE12}, where $\varepsilon_C$ represents the Carnot COP for reversible refrigerators.

The relationship between irreversible thermodynamics and finite-time thermodynamics was first realized by Van den Broeck~\cite{vdbrk2005}. In an elegant investigation on the efficiency at maximum power output for a heat engine, he adopted a generic model shown in Fig.~\ref{fig-engwh}(a) and selected $\dot{Q}_h$, the heat absorbed from the hot bath per unit time, as the thermal flux. Under the assumption of local equilibrium for the heat engine such that the effective temperature of working substance can be well-defined, he found that the efficiency at maximum power output is $\eta_C/2$ up to the linear order for a tight-coupling heat engine within the framework of linear irreversible thermodynamics. In principle, one can also choose $\dot{Q}_c$, the heat released into the cold bath per unit time, as the thermal flux.
Although the arbitrariness in selecting the thermal flux has no serious effect on the major results up to the lowest order for heat engines, this arbitrariness may lead to fatal consequences when we consider some optimization problems for refrigerators. For example, if we follow Van den Broeck's procedure in Ref.~\cite{vdbrk2005} and optimize refrigerators, the results of COP at maximum $\chi-$criterion with taking $\dot{Q}_c$ or $\dot{Q}_h$ as the thermal flux are quite different from each other: One approaches 0 and the other approaches $\sqrt{2\varepsilon_C}$ (which takes a large value) when the temperatures of two baths get close to one another (see details in Appendix~\ref{sec-refg1}). Therefore, it is urgent for us to construct a unified rule to select the thermal flux for a system which is coupled with two heat baths and exchanges work with the outside.

Recent developments in finite-time thermodynamics are focused on the higher order universality~\cite{Schmiedl2008,Tu2008,Esposito2009a,Esposito2009,Espositopre12,Izumida2009,wangx11,Apertet12,Seifert12rev} and the bounds~\cite{Esposito2010,Espositopre2010,GaveauPRL10,WangTu2011,wangtu2012,Izumida2012,WangHe,wangheinter,Izumida2010,Allahverdyan08,Allahverdyan13,Guochen2013} of efficiency at maximum power output for heat engines. It seems that these two issues can hardly be achieved within the conventional generic model shown in Fig.~\ref{fig-engwh}(a).
Instead of this generic model, Esposito \emph{et al.} considered a process of particle transport and verified that the efficiency at maximum power output up to the quadratic order, $\eta_C/2+\eta_C^2/8$, can be achieved for tight coupling between the mass and energy flows and in the presence of a left-right symmetry in the whole system~\cite{Esposito2009,Espositopre12}. In addition, Esposito \emph{et al.} also found the efficiency at maximum power output of heat engines to be bounded between $\eta_C/2$ and $\eta_C/(2-\eta_C)$ under the low-dissipation condition~\cite{Esposito2010}. The similar issues related to the COP at maximum $\chi-$criterion were also addressed in Ref.~\cite{WLTHRpre12,Izumida13,ApertetEPL13,ShengTuJPA13}. It is quite interesting if one can reproduce these results on the basis of the generic model or some revised version.

The above researches raise the following questions: (i) Is there a consistent way to express the generalized thermodynamic fluxes and forces for a heat engine or a refrigerator in contact with two heat baths and outputting or inputting mechanical work within the framework of irreversible thermodynamics? (ii) Can the universal efficiency at maximum power output up to the quadratic order and the bounds of efficiency at maximum power output be achieved on the basis of the generic model of heat engine? In their previous work~\cite{ShengTuJPA13}, the present authors try to solve these two problems by introducing the concept of weighted thermal flux for a system coupling with two heat baths and exchanging work with the outside under the assumption of local equilibrium. Although this work guides a right direction, it still contains some shortcomings. First, we have not exactly expressed the entropy production rate in a canonical form that is the sum of products of generalized thermodynamic fluxes and forces. Second, the physical meaning of the weighted thermal flux is still unclear. Third, we have not explicitly constructed the mappings from two kinds of typical heat engines (such as the low-dissipation heat engine~\cite{Esposito2010} and the Feynman ratchet~\cite{Feynmanbook}) to the generic model. Fourth, as it was commented by Jarzynski~\cite{Jarzynskicom}, the assumption of local equilibrium might be unnecessary because the effective temperature is just a parameter in our previous work~\cite{ShengTuJPA13}.

In this work, we will overcome the above four shortcomings by simultaneously introducing the concepts of weighted reciprocal of temperature and weighted thermal flux. In Sec.~\ref{sec-wtqunt}, we discuss our logic for introducing the concepts of weighted reciprocal of temperature and weighted thermal flux. With the aid of these concepts, the generalized thermodynamic fluxes and forces can be defined in a consistent way without the assumption of local equilibrium for the heat engine.
The entropy production rate can be expressed in a canonical form. The physical meanings of weighted reciprocal of temperature, weighted thermal flux and weighted numbers are clearly interpreted on the basis of a refined generic model. In Sec.~\ref{sec-empheateng}, the universal efficiency at maximum power output up to the quadratic order and the bounds of efficiency at maximum power output are obtained for a tight-coupling engine abiding by a linear constitutive relation between the generalized thermodynamic fluxes and forces. It is surprising that these results share the same forms as those obtained by Esposito \emph{et al.} for the low-dissipation heat engine~\cite{Esposito2010}. In Sec.~\ref{sec-mapping}, we explicitly construct the mappings from two kind of typical heat engines including the low-dissipation heat engine~\cite{Esposito2010} (as a representative of cyclic heat engines) and the Feynman ratchet~\cite{Feynmanbook} (as a representative of autonomous heat engines) into the refined generic model. In particular, using our refined generic model, we find that the low-dissipation heat engine is a kind of tight-coupling engine abiding by a linear constitutive relation. In Sec.~\ref{sec-refrig}, we transplant the concepts of weighted reciprocal of temperature and weighted thermal flux to the optimization of refrigerators. Section \ref{sec-summary} contains a brief summary and discussions.

\section{Generic model and weighted quantities of heat engine\label{sec-wtqunt}}
In this section, we will introduce a generic model and the concepts of weighted reciprocal of temperature and weighted thermal flux for a heat engine.
\subsection{Conventional generic model}
\begin{figure}[htp!]
\includegraphics[width=8cm]{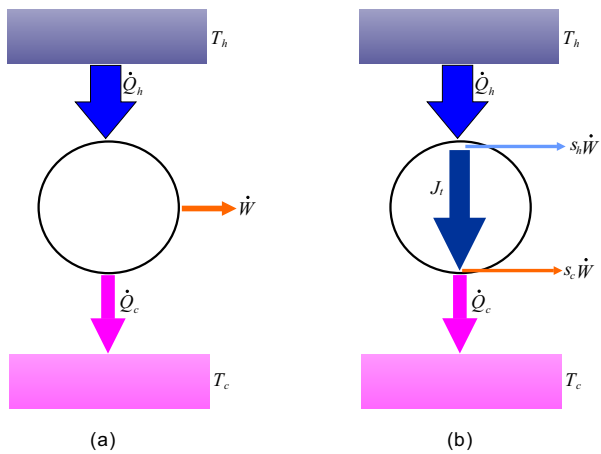}\caption{(Color online) Generic model of a heat engine: (a) the conventional version and (b) the refined version.}\label{fig-engwh}
\end{figure}

Let us consider a generic model of heat engine shown in Fig.~\ref{fig-engwh}(a). The engine absorbs heat $\dot{Q}_h$ per unit time from the hot bath at temperature $T_h$ and outputs power $\dot{W}$. Simultaneously, it releases heat $\dot{Q}_c$ per unit time into the cold bath at temperature $T_c$. For an autonomous engine, we assume it operates in a steady state such that $\dot{Q}_h$, $\dot{Q}_c$ and $\dot{W}$ are time-independent. For a cyclic engine, $\dot{Q}_h$, $\dot{Q}_c$ and $\dot{W}$ represent the average heat ${Q}_h/t_0$ absorbed from the hot bath, the average heat ${Q}_c/t_0$ released into the cold bath, and the average work output $W/t_0$ in each cycle, respectively, where $t_0$ is the period of the cycle.

The first law of thermodynamics requires:
\begin{equation}\dot{Q}_{h}-\dot{Q}_{c}=\dot{W}.\label{eq-Econst}\end{equation}
Since the entropy variation of working substance is vanishing either for a cyclic engine in the whole cycle or an autonomous engine in the steady state, the entropy production rate $\sigma$ may be expressed as~\cite{Prigoginebook}:
\begin{equation}\sigma = \beta _{c}
\dot{Q}_{c}-\beta _{h}\dot{Q}_{h}\label{eq-EnP0}\end{equation}
with $\beta _{c}=1/T_{c}$ and $\beta _{h}=1/T_{h}$. Here and in the following the Boltzmann factor is set to $1$.

\subsection{Weighted reciprocal of temperature and weighted thermal flux}
Now we will try to express the entropy production rate in a canonical form, i.e., the sum of products of generalized thermodynamic fluxes and forces within the framework of irreversible thermodynamics. On the one hand, equation~(\ref{eq-Econst}) implies $\dot{Q}_{c}=\dot{Q}_{h}-\dot{W}$, from which Eq.~(\ref{eq-EnP0}) can be transformed into
\begin{equation}\sigma =-\beta _{c}\dot{W}+\dot{Q}_{h}\left( \beta _{c}-\beta _{h}\right).\label{eq-EnP1}\end{equation}
On the other hand, equation~(\ref{eq-Econst}) implies $\dot{Q}_{h}=\dot{Q}_{c}+\dot{W}$, from which Eq.~(\ref{eq-EnP0}) can be transformed into
\begin{equation}\sigma =-\beta _{h}\dot{W}+\dot{Q}_{c}\left( \beta _{c}-\beta _{h}\right).\label{eq-EnP2}\end{equation}

Introduce two nonnegative weighted numbers $s_c$ and $s_h$ such that $s_c+s_h =1$. Multiply Eqs.~(\ref{eq-EnP1}) and (\ref{eq-EnP2}) by $s_c$ and $s_h$, respectively. Then the sum of both products leads to
\begin{equation}\sigma =-(s_{c}\beta _{c}+s_{h}\beta _{h}) \dot{W}+(s_{c}\dot{Q}_{h}+ s_{h}\dot{Q}_{c}) (\beta _{c}-\beta_{h}), \label{eq-EnPf}\end{equation}
which enlightens us to define the weighted reciprocal of temperature
\begin{equation}\beta\equiv s_{c}\beta _{c}+s_{h}\beta _{h}\label{eq-wttemp}\end{equation}
and weighted thermal flux
\begin{equation}J_t\equiv  s_{c}\dot{Q}_{h}+ s_{h}\dot{Q}_{c}.\label{eq-wtthflux}\end{equation}
The generalized force conjugated to $J_t$ may be defined as
\begin{equation}X_t\equiv \beta _{c}-\beta_{h}.\label{eq-wtthforce}\end{equation}
The generalized thermodynamic flux and force related to the mechanical process, which are briefly called the mechanical flux and force, can be expressed as
\begin{equation}J_m\equiv  1/t_0,\mathrm{~and~} X_m\equiv -\beta W,\label{eq-mechfluxforce}\end{equation}
respectively for the cyclic engine.
For the autonomous engine operating in the steady state, the power output $\dot{W}$ can be expressed as the product of the net rate $r$ and the elementary work $w$ in each mechanical step. The mechanical flux and force may be expressed as
\begin{equation}J_m\equiv  r,\mathrm{~and~} X_m\equiv  -\beta w,\label{eq-mechfluxforce1}\end{equation}
respectively for the autonomous engine. The minus sign is added in the expression of $X_m$ because the heat engine performs output work rather than input work.

We emphasize that the analogous form of $J_{m}$ and $X_{m}$ in Eq.(\ref{eq-mechfluxforce}) was first adopted by Izumida \emph{et al.}~\cite{Izumida2010,Izumida2012} for the cyclic engine. In comparison with the previous work~\cite{vdbrk2005}, this kind of definitions of the mechanical flux and force for the cyclic engine has three advantages: (i) it is not necessary to introduce the effective temperature of working substance when we define the generalized thermodynamic force related to the external load (briefly called the mechanical force in the present paper). (ii) The dimensions of mechanical flux and force for cyclic engine are the same as those for autonomous engine, respectively. (iii) With the consideration of Eqs.~(\ref{eq-wttemp})--(\ref{eq-mechfluxforce1}), we can exactly transform Eq.~(\ref{eq-EnPf}), the entropy production rate, into a canonical form
\begin{equation}\sigma =J_m X_m +J_t X_t.\label{eq-entcanon}\end{equation}

\subsection{Physical meanings of the weighted quantities and refined generic model of heat engine\label{sec-physmean}}

In the above definitions of generalized thermodynamic fluxes and forces, the assumption of local equilibrium for the heat engine, based on which the effective temperature of working substance is well defined, is discarded through introducing the concept of weighted reciprocal of temperature (\ref{eq-wttemp}). If we still adopt the assumption of local equilibrium, then $\beta^{-1}$ might be regarded as an effective temperature of working substance. It is beyond the scope of this paper to construct an example of statistical mechanics to confirm this interpretation. However
our following discussions are still meaningful because we just treat $\beta^{-1}$ as a parameter defined by Eq.~(\ref{eq-wttemp}) rather than the effective temperature in the following content.

In addition, from Eqs.~(\ref{eq-Econst}) and (\ref{eq-wtthflux}), we can derive
\begin{equation}\dot{Q}_{h}=J_{t}+s_{h}\dot{W},~\dot{Q}_{c}=J_{t}-s_{c}\dot{W},\label{eq-meanings}\end{equation}
from which one may revise the conventional generic model in Fig~\ref{fig-engwh}(a) into a refined version in Fig~\ref{fig-engwh}(b). The engine absorbs heat $\dot{Q}_h$ per unit time from the hot bath, an amount of heat $s_h \dot{W}$ will be transformed into the work output per unit time due to the coupling between the engine and the hot bath. A thermal flux $J_t$ flows through the heat engine, then an amount of heat $s_c \dot{W}$ will be transformed into the work output per unit time due to the coupling between the engine and the cold bath. Finally, the engine releases heat $\dot{Q}_c$ per unit time into the cold bath.
Based on this picture, we can directly write out the entropy production rate in canonical form (\ref{eq-entcanon}) according to Eq.~(3.12) in the book by Prigogine~\cite{Prigoginebook} if we consider the extended system including two heat baths and the engine (the details are shown in Appendix~\ref{sec-Prig}).

With the consideration of Eq.~(\ref{eq-mechfluxforce}) or (\ref{eq-mechfluxforce1}), the power output can be expressed
\begin{equation}\dot{W}=-\beta^{-1}J_m X_m.\label{eq-power1}\end{equation}
Since the mechanical flux is in the liner order for small thermodynamic force, Eq.~(\ref{eq-power1}) implies that the leading term of $\dot{W}$ is a quadratic order for small thermodynamic forces.
$\dot{Q}_{h}$ should at least contain the linear term since $\dot{W}/\dot{Q}_{h}\le \eta_C \rightarrow 0$ for small relative temperature.
Combining Eq.~(\ref{eq-meanings}), we regard $J_t$ as the common leading term shared by $\dot{Q}_{c}$ and $\dot{Q}_{h}$. The parameter $s_c$ (or $s_h$) represents the fraction of power output occupied by the higher order term in $\dot{Q}_{c}$ (or $\dot{Q}_{h}$). We argue that $s_c$ (or $s_h$) depends on the degree of coupling between the engine and the cold (or hot) bath. That is, the coupling between the working substance and the baths is significant for finite-time heat engines while it is unimportant for the conventional quasi-static heat engines.

\section{Efficiency at maximum power output for a tight-coupling heat engine\label{sec-empheateng}}
Maximizing the power output $\dot{W}$ in Eq.~(\ref{eq-power1}) with respect to $X_m$ for given $T_c$ and $T_h$, we obtain
\begin{equation}
X_m(\partial J_m/\partial X_m)+J_m =0.\label{eq-maxp0}\end{equation}
Let us consider a tight-coupling heat engine, in which the weighted thermal flux is proportional to the mechanical flux, satisfying
\begin{equation}
J_t/J_m = \xi,\label{eq-tightc0}\end{equation}
where $\xi$ is a quantity independent of the thermodynamic forces. Then the efficiency $\eta\equiv \dot{W}/Q_h$ is transformed into
\begin{equation}\eta=\frac{-X_m/\beta\xi}{1-s_hX_m/\beta\xi}\label{eq-eta0}\end{equation} with the consideration of Eqs.~(\ref{eq-meanings}) and (\ref{eq-power1}).

For simplicity, we consider a linear constitutive relation between the thermodynamic fluxes and forces:
\begin{equation}J_t=L_{tt}X_t +L_{tm}X_m,~J_m=L_{mt}X_t +L_{mm}X_m,\label{eq-linearcons}\end{equation}
where the Onsager coefficients satisfy $L_{mt}=L_{tm}$, $L_{tt}\ge 0$, $L_{mm}\ge0$, and $L_{tt}L_{mm}\ge L_{mt}L_{tm}$. Combining tight-coupling condition (\ref{eq-tightc0}) with linear constitutive relation (\ref{eq-linearcons}), we obtain
\begin{equation}L_{tt}/L_{tm}=L_{mt}/L_{mm}=\xi.\label{eq-linearcons2}\end{equation}
Substituting Eq.~(\ref{eq-linearcons}) into Eq.~(\ref{eq-maxp0}), we can derive the optimal mechanical force
\begin{equation}X_m^{\ast}=-(\xi/2)X_t \label{eq-optXm}\end{equation}
with the consideration of Eq.~(\ref{eq-linearcons2}). Substituting the above equation into Eq.~(\ref{eq-eta0}), we obtain the efficiency at maximum power output
\begin{equation}\eta^\ast= \frac{\eta _{C}}{2-s_{h}\eta _{C}}\label{eq-etamaxp}\end{equation}
with the consideration of $\beta_c\equiv 1/T_c$, $\beta_h\equiv 1/T_h$, $\eta _{C}\equiv 1-T_c/T_h$, Eqs.~(\ref{eq-wttemp}) and (\ref{eq-wtthforce}). This result shares the same form as the efficiency at maximum power output for the low-dissipation heat engine~\cite{Esposito2010} (or the stochastic heat engine~\cite{Schmiedl2008} which can be regarded as a special low-dissipation heat engine).

When the heat engine couples symmetrically with the cold and the hot baths, $s_c=s_h=1/2$, equation~(\ref{eq-etamaxp}) is transformed into $\eta^\ast=2\eta _{C}/(4-2\eta _{C})\approx \eta_{C}/2+\eta_{C}^2/8+\cdots$. This fact is consistent with the universal efficiency at maximum power output ($\eta_{C}/2+\eta_{C}^2/8$) up to the quadratic order for tight-coupling heat engines in the presence of left-right symmetry~\cite{Esposito2009,Espositopre12}. On the other hand, equation~(\ref{eq-etamaxp}) implies that $\eta^\ast$ increases monotonically with $s_h$. It follows that $\eta^\ast$ is bounded between the lower bound $\eta_{-}\equiv\eta_C/2$ and the upper one $\eta_{+}\equiv \eta _{C}/(2-\eta _{C})$. Furthermore, the lower and upper bounds are reached in the case of extremely asymmetric coupling. These bounds are also shared by the low-dissipation engine and reached in the case of extremely asymmetric dissipations~\cite{Esposito2010}. These surprising results enlighten us to investigate the relationship between our refined generic model and the low-dissipation engine.

\section{Mapping two kinds of typical heat engines into our refined generic model\label{sec-mapping}}

In this section, we will construct the mappings from two kinds of typical heat engines such as the low-dissipation engine and the Feynman ratchet into our refined generic model.
\subsection{Low-dissipation engine\label{sec-lowdispat}}
A low-dissipation engine~\cite{Esposito2010} undergoes a thermodynamic cycle consisting of two ``isothermal" and two adiabatic processes. The word ``isothermal" merely indicates that the heat engine is in contact with a heat bath at constant temperature.
In the process of ``isothermal" expansion during time interval $t_h$, the engine absorbs heat $Q_h$ from the hot bath at temperature $T_h$. The variation of entropy in this process is denoted as $\Delta S$. On the contrary, In the process of ``isothermal" compression during time interval $t_c$, the engine releases heat $Q_c$ into the cold bath at temperature $T_c$. There is no heat exchange and entropy production in two adiabatic processes. Assume that the time for completing the adiabatic processes is negligible relative to $t_c$ and $t_h$. So the period of the whole cycle is $t_0=t_c+t_h$. The entropy production in each  ``isothermal" process is assumed to be proportional to the reciprocal of time interval for completing that process, which is called low-dissipation assumption~\cite{Esposito2010}. This assumption is quite reasonable for large enough $t_0$. According to this assumption, the heats $Q_h$ and $Q_c$ can be expressed as
\begin{equation}Q_{h} =T_{h}( \Delta S-\Gamma_{h}/t_{h}),~
-Q_{c} =T_{c}( -\Delta S-\Gamma_{c}/t_{c}),\label{eq-lowdisassp}\end{equation}
with two dissipation coefficients $\Gamma_{h}$ and $\Gamma_{c}$, respectively.

Introducing two parameters $\alpha_c\equiv t_c/t_0$ and $\alpha_h\equiv t_h/t_0$ which satisfy $\alpha_c+\alpha_h=1$, we obtain $\dot{Q}_h\equiv{Q}_h/t_0=T_{h}\Delta S/t_{0}-T_{h}\Gamma _{h}/\alpha
_{h}t_{0}^{2}$ and $\dot{Q}_c\equiv{Q}_c/t_0=T_{c}\Delta S/t_{0}+T_{c}\Gamma _{c}/\alpha
_{c}t_{0}^{2}$. The work output $W\equiv Q_{h}-Q_{c}$ can be expressed as
\begin{equation}W =( T_{h}-T_{c}) \Delta S-( T_{h}\Gamma
_{h}/\alpha _{h}+T_{c}\Gamma _{c}/\alpha _{c}) /t_{0}.\label{eq-workoutp}\end{equation}
The weighted thermal flux (\ref{eq-wtthflux}) can be further expressed as
$J_{t}=(s_{c}T_{h}+s_{h}T_{c})\Delta S/t_{0}+( s_{h}T_{c}\Gamma _{c}/\alpha _{c}-s_{c}T_{h}\Gamma
_{h}/\alpha _{h}) /t_{0}^{2}$. Because $J_{t}$ is regarded as the common leading term of $\dot{Q}_{c}$ and $\dot{Q}_{h}$ as mentioned in Sec.IIC, we require $s_{h}T_{c}\Gamma_{c}/\alpha _{c}-s_{c}T_{h}\Gamma
_{h}/\alpha _{h}=0$ to keep the quadratic term vanishing. With consideration of $s_c+s_h=1$, the weighted parameters may be expressed as
\begin{equation}s_{h}=\frac{T_{h}\Gamma _{h}/\alpha _{h}}{T_{h}\Gamma
_{h}/\alpha _{h}+T_{c}\Gamma _{c}/\alpha _{c}},~s_{c}=\frac{T_{c}\Gamma
_{c}/\alpha _{c}}{T_{h}\Gamma _{h}/\alpha _{h}+T_{c}\Gamma _{c}/\alpha _{c}}.\label{eq-cpparm}\end{equation}
It is easy to see that these weighted parameters indeed reflect the degree of coupling between the heat engine and the hot bath or the cold one.
With these weighted parameters, the weighted reciprocal of temperature and the weighted thermal flux can be expressed as
\begin{equation}\beta=\frac{\Gamma
_{h}/\alpha _{h}+\Gamma _{c}/\alpha _{c}}{T_{h}\Gamma _{h}/\alpha _{h}+T_{c}\Gamma _{c}/\alpha _{c}},\label{eq-betaloweng}\end{equation}
and
\begin{equation}J_t=\beta T_{h}T_{c}\Delta S/t_{0},\label{eq-Jtloweng}\end{equation}
respectively. The above equation implies that the thermal flux $J_t$ couples tightly with the mechanical flux $J_m\equiv 1/t_0$. In this sense, the low-dissipation heat engine is exactly a kind of tight-coupling heat engine.

Considering definitions (\ref{eq-wtthforce}) and (\ref{eq-mechfluxforce}), we derive
\begin{equation}
J_m =\frac{X_m}{\Gamma _{h}/\alpha
_{h}+\Gamma _{c}/\alpha _{c}}+\frac{T_{h}T_{c}\Delta S X_t}{T_{h}\Gamma _{h}/\alpha
_{h}+T_{c}\Gamma _{c}/\alpha _{c}},
\label{eq-Jmloweng}\end{equation}
from Eqs.~(\ref{eq-workoutp}) and (\ref{eq-betaloweng}). With consideration of Eqs.~(\ref{eq-linearcons}) and (\ref{eq-betaloweng})--(\ref{eq-Jmloweng}), we obtain
the Onsager coefficients
\begin{eqnarray}
&&L_{mm}=\frac{1}{\Gamma_{h}/\alpha_{h}+\Gamma _{c}/\alpha _{c}},\nonumber \\
&&L_{mt}=L_{tm}=\frac{T_{h}T_{c}\Delta S}{T_{h}\Gamma _{h}/\alpha_{h}+T_{c}\Gamma_{c}/\alpha_{c}}, \label{eq-OnsagerLEG}\\
&&L_{tt}=\frac{( \Gamma _{h}/\alpha _{h}+\Gamma _{c}/\alpha
_{c}) (T_{h} T_{c}\Delta S) ^{2}}{( T_{h}\Gamma
_{h}/\alpha _{h}+T_{c}\Gamma _{c}/\alpha _{c})^{2}}.\nonumber
\end{eqnarray}
Thus the low-dissipation engine can be strictly mapped into our refined generic model shown in Fig.~\ref{fig-engwh}(b) with the Onsager coefficients and weighted parameters satisfying Eqs.~(\ref{eq-OnsagerLEG}) and (\ref{eq-cpparm}), respectively. Similar procedure is also available for a tight-coupling thermoelectric generator investigated in Ref.~\cite{ApertetPRE13}, which is shown in Appendix~\ref{sec-thermEG}.

With consideration of Eqs.~(\ref{eq-power1}), (\ref{eq-linearcons2}), (\ref{eq-optXm}), (\ref{eq-betaloweng}) and (\ref{eq-OnsagerLEG}), the power output after being optimized with respect to $X_m$ can be expressed as
\begin{equation}\dot{W}=\frac{X_{t}^{2}}{4}\frac{( T_{c}T_{h}\Delta S)^{2}}{(
T_{h}\Gamma _{h}/\alpha _{h}+T_{c}\Gamma_{c}/\alpha _{c})}.\end{equation}
Since $\alpha _{h}+\alpha _{c}=1$, the above expression takes maximum when $\alpha _{h}/\alpha _{c}=\sqrt{T_{h}\Gamma _{h}/T_{c}\Gamma _{c}}$. Thus we can derive \begin{equation}s_h={\sqrt{T_{h}\Gamma _{h}}}/{(\sqrt{T_{h}\Gamma _{h}}+\sqrt{T_{c}\Gamma _{c}})}\label{eq-sh2}\end{equation} from Eq.~(\ref{eq-cpparm}). Finally, from Eq.~(\ref{eq-etamaxp}) the efficiency at maximum power output for the low-dissipation engine can be expressed as
\begin{equation}\eta^\ast=\frac{\eta _{C}}{2-{\sqrt{T_{h}\Gamma _{h}}\eta _{C}}/{(\sqrt{T_{h}\Gamma _{h}}+\sqrt{T_{c}\Gamma _{c}})}},\end{equation}
which is the same as the result obtained by Esposito \emph{et al.} for the low-dissipation engine~\cite{Esposito2010}.

\subsection{Feynman ratchet}
\begin{figure}[htp!]\begin{center}
\includegraphics[width=7cm]{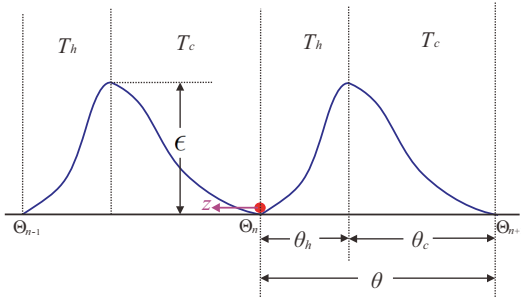}\caption{\label{fig-Feyman}(Color online) Schematic digram of Feynman ratchet as a heat engine.}\end{center}
\end{figure}

The Feynman ratchet~\cite{Feynmanbook} can be regarded as the B\"uttiker-Landauer model~\cite{Buttiker,Landauer}, i.e., a Brownian particle walking in a periodic lattice labeled by $\Theta_n,~(n=\cdots,-2,-1,0,1,2,\cdots)$ with a fixed step size $\theta$. The ratchet potential is schematically depicted in Fig.~\ref{fig-Feyman}, where the energy scale and the position of potential barrier are respectively denoted by $\epsilon$ and $\theta_h$. The Brownian particle is in contact with a hot bath at temperature $T_h$ in the left side of each potential barrier while it is in contact with a cold bath at temperature $T_c$ in the right side of each barrier. The particle moves against a load $z$ and outputs work. In the steady state and under the overdamping condition, the forward and backward jumping rates can be respectively expressed as $R_{F} =r_{0}\mathrm{e}^{-\beta_{h}(\epsilon +z\theta _{h})}$ and $R_{B}=r_{0}\mathrm{e}^{-\beta _{c}(\epsilon -z\theta _{c})}$ according to the Arrhenius law \cite{Feynmanbook} with $\beta_{h}=1/T_h$, $\beta_{c}=1/T_c$, and $\theta _{c}=\theta-\theta _{h}$. Here $r_{0}$ represents the bare rate constant with dimension of time$^{-1}$.

The net rate may be defined as
\begin{equation}r\equiv R_{F}-R_{B}=r_{0}\left[ \mathrm{e}^{-\beta _{h}( \epsilon +z\theta _{h})
}-\mathrm{e}^{-\beta _{c}( \epsilon -z\theta _{c}) }\right].\label{eq-fluxFeym}\end{equation}
If we consider only the energy transaction due to the potential energy, in each forward step, the particle absorbs heat $q_h\equiv \epsilon+z{\theta_h}$ from the hot bath. After doing output work $w\equiv z\theta$ against the external load, the remained heat $q_c\equiv q_h-w=\epsilon -z\theta _{c}$ will be released into the cold bath when the particle jumps over the barrier. The energy conversion in each backward step is exactly opposite of that in the forward step mentioned above.
With consideration of the net rate $r$, the net power output can be expressed as
\begin{equation}\dot{W}=z\theta r\label{eq-network}, \end{equation}
while the heat absorbed from the hot bath and the heat released into the cold bath per unit time can be respectively expressed as
\begin{equation}\dot{Q}_{h}=(\epsilon+z\theta _{h})r=\epsilon r+ (\theta _{h}/\theta)\dot{W}\label{eq-Qhot}, \end{equation}
and
\begin{equation}\dot{Q}_{c}=(\epsilon -z\theta _{c})r=\epsilon r- (\theta _{c}/\theta)\dot{W}\label{eq-Qcold}. \end{equation}
Considering the above two equations and Eq.~(\ref{eq-meanings}), we can straightforwardly write out the weighted thermal flux
\begin{equation}J_t=\epsilon r\label{eq-JtFeym},\end{equation}
and the weighted parameters
\begin{equation}s_h=\theta _{h}/\theta,~s_c=\theta _{c}/\theta\label{eq-wtnumFeym}.\end{equation}
Obviously, the above weighted parameters reflect the asymmetric degree of the ratchet potential in Fig.~\ref{fig-Feyman} which is equivalent to the asymmetric degree of the coupling regions between the Brownian particle and the hot bath or the cold one.

On the other hand, according to Eq.~(\ref{eq-mechfluxforce1}), the mechanical flux and force can be expressed as
\begin{equation}X_m=-\beta z\theta,~J_m =r\label{eq-JmFeym}\end{equation}
where $\beta$ and $r$ satisfy Eqs.~(\ref{eq-wttemp}) and (\ref{eq-fluxFeym}), respectively. Thus Eqs.~(\ref{eq-JtFeym}) and (\ref{eq-JmFeym}) imply that the thermal flux couples tightly with the mechanical flux. This fact holds because we have merely considered the energy transaction due to the potential energy. It might be broken if the kinetic energy cannot be neglected~\cite{Parrondo96,Sekimoto97,Hondou2000}. The specific discussions on this point are beyond the scope of our present work.

So far, the Feynman ratchet can be strictly mapped into our generic model shown in Fig.~\ref{fig-engwh}(b). Under the linear approximation, equation~(\ref{eq-fluxFeym}) can be further expressed as $r=r_{0}\mathrm{e}^{-\beta \epsilon }(X_{m}+\epsilon X_{t})$. Substituting it into Eqs.~(\ref{eq-JtFeym}) and (\ref{eq-JmFeym}), and comparing the results with Eq.~(\ref{eq-linearcons}), we can derive the Onsager coefficients
\begin{equation}L_{mm} =r_{0}\mathrm{e}^{-\beta \epsilon },~L_{mt}=L_{tm}=r_{0}\mathrm{e}^{-\beta
\epsilon }\epsilon,~L_{tt}=r_{0}\mathrm{e}^{-\beta
\epsilon }\epsilon ^{2}.\label{eq-OnsagerFeym}\end{equation}
Therefore, the Feynman ratchet can only be mapped into an approximately linear model with the Onsager coefficients and weighted parameters satisfying Eqs.~(\ref{eq-OnsagerFeym}) and (\ref{eq-wtnumFeym}), respectively. The direct consequence is that the universality and bounds of efficiency at maximum power output derived on the basis of the linear constitutive relation (\ref{eq-linearcons}) in Sec.~\ref{sec-empheateng} might not hold if we consider the nonlinear terms.

\subsection{Universality of efficiency at maximum power output for heat engines abiding by a nonlinear constitutive relation\label{sec-unvEMP}}
If the constitutive relation contains nonlinear terms, the mechanical flux for a tight-coupling engine can be formally expressed as
\begin{equation}J_m=L_{mm}(X_m+\xi X_t)+C_{m}X_m^2+C_{t}X_t^2+C_{mt}X_mX_t+O(X_m^3,X_t^3). \label{eq-Jmgeneral}\end{equation}
This equation is consistent with Eqs.~(\ref{eq-linearcons}) and (\ref{eq-linearcons2}) under the linear approximation. $O(X_m^3,X_t^3)$ represents the third and higher order terms.
The coefficients $C_{m}$, $C_{t}$ and $C_{mt}$ of the quadratic order terms in Eq.~(\ref{eq-Jmgeneral}) may depend on $T_c$, $T_h$, $\xi$, $s_c$, and $s_h$.

In the case of symmetric coupling ($s_h=s_c=1/2$), the thermodynamic fluxes should be exactly reversed when the thermodynamic forces are reversed; that is, $J_m\mapsto -J_m$ under the operation $X_m\mapsto -X_m$ and $X_t\mapsto -X_t$, which requires that the coefficients $C_{m}$, $C_{t}$ and $C_{mt}$ of the quadratic order terms in Eq.~(\ref{eq-Jmgeneral}) are vanishing. This fact can be confirmed if we expand Eq.~(\ref{eq-fluxFeym}) up to the higher order (see Appendix~\ref{sec-FeynH}). It can also be verified with the maser model in Ref.~\cite{Esposito2009}. Through simple calculations, we find that the quadratic order terms of the current, Eq.~(23) in Ref.~\cite{Esposito2009}, is indeed vanishing for the symmetric coupling $\Gamma_l=\Gamma_r$.

Substituting Eq.~(\ref{eq-Jmgeneral}) with vanishing $C_{m}$, $C_{t}$ and $C_{mt}$ into Eq.~(\ref{eq-maxp0}), we obtain the optimal mechanical force:
\begin{equation}X_m^{\ast}=-(\xi/2)X_t +O(X_t^3).
\label{eq-optXm2}\end{equation}
It is not hard to see that the third order term $O(X_t^3)$ has no effect on the quadratic term in the expression of efficiency at maximum power output when we substitute Eq.~(\ref{eq-optXm2}) into Eq.~(\ref{eq-eta0}). Thus the universal efficiency at maximum power output ($\eta_C/2+\eta_C^2/8$) holds exactly up to the quadratic order when the heat engine couples symmetrically and tightly with two baths.

\section{Transplanting the weighted reciprocal of temperature and weighted thermal flux to the optimization of refrigerators\label{sec-refrig}}
It is straightforward to transplant the similar procedure to the optimization of refrigerators. All arrows in Fig.~\ref{fig-engwh} are reversed for a refrigerator. The entropy production rate can be expressed as $\sigma =\beta _{h}\dot{Q}_{h} -\beta _{c}
\dot{Q}_{c}$. Through simple calculations, we find that the weighted reciprocal of temperature, the weighted thermal flux and the mechanical flux have the same forms as those in Sec.~\ref{sec-wtqunt}, while the generalized thermal force and mechanical force take the opposite sign for the refrigerator. In other words, equations (\ref{eq-wttemp})--(\ref{eq-entcanon}) still hold except for $X_t\equiv \beta_h -\beta_c$ and $X_m \equiv \beta W$ (or $\beta w$). The details are shown in Appendix~\ref{sec-appxe}.

The power input can be expressed as $\dot{W}=\beta^{-1} J_m X_m$ for the refrigerator.
The heat absorbed from the cold bath can be expressed as
\begin{equation}
\dot{Q}_{c}=J_{t}-s_{c}\dot{W}=(\xi -s_{c}\beta^{-1} X_m)J_m \label{eq-heatabref}\end{equation} with consideration of the tight-coupling condition (\ref{eq-tightc0}). The COP can be further expressed as
\begin{equation}\varepsilon \equiv \dot{Q}_{c}/\dot{W}= \xi\beta/X_m -s_c.\label{eq-coprefig}\end{equation}
Thus the target function $\chi\equiv \varepsilon \dot{Q}_c$, suggested in
Refs.~\cite{YanChen1990} and \cite{RocoPRE12} for the refrigerator, can be transformed into
\begin{equation}\chi=(\xi-s_cX_m /\beta)^2\beta J_m /X_m.\end{equation}
Maximizing $\chi$ with respect to $X_m$ for given $T_c$ and $T_h$, we obtain
\begin{equation}\label{eq-xmmaxchi}
\left(s_c+\frac{\beta \xi}{X_m}\right) \frac{J_m}{X_m} +\left(s_c -\frac{\beta \xi}{X_m}\right)\frac{\partial J_m}{\partial X_m}=0.
\end{equation}

Now we consider the linear constitutive relation that satisfies Eqs.~(\ref{eq-linearcons}) and (\ref{eq-linearcons2}). From Eq.~(\ref{eq-xmmaxchi}), we can derive
\begin{equation}{\beta \xi}/{ X_m} = \sqrt{{s_c^2}/{4}-{2s_c\beta}/{X_t}}-{s_c}/{2}.\end{equation}
Substituting this equation into Eq.~(\ref{eq-coprefig}), we obtain the COP at maximum $\chi$:
\begin{equation}\varepsilon _{\ast} =\sqrt{2s_{c}\varepsilon _{C}+9s_{c}^{2}/4}-3s_{c}/2\label{eq-copMaxchi}\end{equation}
with the consideration of $X_t\equiv\beta_h -\beta_c$, $\varepsilon_C\equiv T_c/(T_h-T_c)$, and Eq.~(\ref{eq-wttemp}).
Obviously, in the case of symmetric coupling, $s_c=s_h=1/2$, the above equation leads to
\begin{equation}\varepsilon _{\ast} =\sqrt{\varepsilon _{C}+9/16}-3/4 \label{eq-copMaxchi2}\end{equation}
whose leading term is $\sqrt{\varepsilon _{C}}$ for small relative temperature difference $\varepsilon_C^{-1}\equiv (T_h-T_c)/T_c \ll 1$. In addition, we can prove that the COP at maximum $\chi$ is bounded between 0 and $(\sqrt{8\varepsilon _{C}+9}-3) /2$. These bounds are reached for extremely asymmetric coupling $s_{c}=0$ and 1, respectively. It is easy to see that these bounds are the same as those for low-dissipation refrigerators~\cite{WLTHRpre12}.

By analogy with the discussion on the symmetrically tight-coupling heat engine in Sec.~\ref{sec-unvEMP}, we can also prove that the universal COP at maximum $\chi$, $\varepsilon _{\ast} \simeq\sqrt{\varepsilon _{C}}$, still holds for a symmetrically tight-coupling refrigerator abiding by a nonlinear constitutive relation.

\section{Conclusion and discussion\label{sec-summary}}

In summary, we have introduced the concepts of weighted reciprocal of temperature and weighted thermal flux
for a generic model of heat engines and recovered many key results on the efficiency at maximum power output for heat engines. We also transplant these concepts to a generic model of refrigerators. The mappings from two kinds of typical heat engines such as the low-dissipation engine and the Feynman ratchet into our refined generic model are constructed. We find that the universality of efficiency at maximum power output up to the quadratic order holds for a heat engine symmetrically and tightly coupled with two baths regardless of linear or nonlinear constitutive relation.

Before ending this paper, we address some issues which have not been fully touched in the main text.

i). We emphasize that the concept of weighted thermal flux has been introduced under the assumption of local equilibrium in our previous work~\cite{ShengTuJPA13}. However, the entropy production rate cannot be exactly expressed in canonical form (\ref{eq-entcanon}). Only within the present framework where the weighted reciprocal of temperature and the weighted thermal flux are simultaneously introduced, we can rigorously express the entropy production rate in canonical form (\ref{eq-entcanon}) and then map two kinds of typical heat engines such as the low-dissipation engine and the Feynman ratchet into our refined generic model.

ii). If we merely keep the thermodynamic fluxes $J_m$ and $J_t$ up to the linear order, then Eq.~(\ref{eq-meanings}) tells us that $\dot{Q}_h$ and $\dot{Q}_c$ still contain quadratic terms $s_h\dot{W}$ and $s_c\dot{W}$, respectively. This property is similar to the minimally nonlinear model proposed by Izumida and Okuda~\cite{Izumida2012} where the minimal nonlinear terms $\gamma_hJ^2_m$ and $\gamma_cJ^2_m$ are intuitively imposed. These minimal nonlinear terms are interpreted as the inevitable power
dissipations accompanied by the finite-time motion of the
heat engines~\cite{Izumida2012}. However, the nonlinear terms in our model are mathematically strict, and they naturally enter into the expressions of $\dot{Q}_h$ and $\dot{Q}_c$. In addition, the Onsager coefficients in Eq.~(\ref{eq-OnsagerLEG}) are slightly different from those obtained in Ref.~\cite{Izumida2012}. After a careful analysis, we find that this distinction comes from different definitions of thermal flux and mechanical force in these two models. The minimally nonlinear irreversible heat engine can be mapped into our refined generic model as shown in Appendix~\ref{sec-appxg}.

iii). Our proof to the universality of efficiency at maximum power output up to the quadratic order is quite different from the procedure of Esposito \emph{et al.} who adopted a general model system of particle transport~\cite{Esposito2009,Espositopre12}. They found that the universality of efficiency at maximum power output up to the quadratic order holds for tight coupling between the mass and energy flows and in the presence of a left-right symmetry in the whole system. However, it is still a challenge for us to connect our proof and theirs at the present stage.

iv). Our discussion implies that the symmetric coupling is a sufficient condition for validating the universality of efficiency at maximum power output up to the quadratic order. However, it is not a necessary condition. For example, if we consider the quasi-symmetric coupling $s_h =1/2 +O(\eta_C)$ where $O(\eta_C)$ represents the linear order of $\eta_C$, we can still derive $\eta^\ast=\eta_C/2+\eta_C^2/8+O(\eta_C^3)$ from Eq.~(\ref{eq-etamaxp}). In addition, as shown in Appendix~\ref{sec-appxf}, the efficiency at maximum power output up to the quadratic order is still $\eta_C/2+\eta_C^2/8$ for the Curzon-Ahlborn endoreversible heat engine although this engine is not a strictly tight-coupling engine. The universality is even independent of the symmetry of couplings between the Curzon-Ahlborn heat engine and two heat baths. It is relative difficult for us to find a sufficient and necessary condition for validating the universality of efficiency at maximum power output up to the quadratic order.

v). We have verified that the efficiency at maximum power output is bounded between $\eta_C/2$ and $\eta_C/(2-\eta_C)$ for a tight-coupling heat engine which abides by the linear constitutive relation. We do not know to what extend these bounds are still valid for a tight-coupling heat engine which abides by the nonlinear constitutive relation.

vi). Besides concerning finite-time operation, finite-time thermodynamics also concerns friction. Bizarro \emph{et al.} investigated the effect of friction on the entropy production and efficiency of heat engines~\cite{Bizarro2010JAP,Bizarro2012AJP,Bizarro2012PRE}. We note that he also derived a form of weighted reciprocal of temperature, which is analogous to Eq.~(\ref{eq-wttemp}), when he discussed the rate of entropy production due to friction in the interface between the heat engine and the heat bath. However, the weighted parameters defined in his model and ours look quite different. The weighted parameters in Ref.~\cite{Bizarro2010JAP} represent the fraction of frictional work dissipated into the engine (or the bath) in the total frictional work, while the weighted parameters in the present work represent the degree of coupling between the engine and the cold (or hot) bath. Although we have not found the explicit connection between these two models yet, the similarity of weighted reciprocal of temperatures in these two models suggests that it is possible to generalize our refined generic model with consideration of the friction.

The above issues are very significant to both irreversible thermodynamics and finite-time thermodynamics. We will make efforts to overcome the difficult problems mentioned above in future research.

\section*{Acknowledgement}
The authors are grateful to financial supports from the National Natural Science Foundation of China (Grant Nos. 11322543 and 11075015). ZCT also thanks Christopher Jarzynski, Massimiliano Esposito, Haitao Quan and Yang Wang for their instructive discussions and suggestions.

\appendix
\section{Optimization of refrigerators following Van den Broeck's procedure\label{sec-refg1}}
As was done by Van den Broeck~\cite{vdbrk2005} for heat engines, here we consider a generic setup for a tight-coupling refrigerator. An external force $F$ is applied on the system and inputs a power $P=F\dot{x}$ into the system, where $x$ is the thermodynamically conjugate variable of $F$. The dot represents the derivative with respect to time. The corresponding thermodynamic force may be taken as $X_{1}={F}/{T}$, where $T$ is the
temperature of the refrigerator which can be well defined due to the assumption of local equilibrium. The thermodynamic flux conjugated to $X_{1}$ is $J_1=\dot{x}$. Then the power input can be expressed as $P=J_{1}X_{1}T$ in terms of the mechanical flux and force.

Let $J_2$ and $X_2\equiv 1/T_h -1/T_c$ denote the generalized thermal flux and force, respectively. According to linear irreversible thermodynamics, we write the linear constitutive relation between the thermodynamic fluxes and forces:
\begin{equation}J_{1}=L_{11}X_{1}+L_{12}X_{2},~  J_{2}=L_{21}X_{1}+L_{22}X_{2},\label{eq-lirrevt}\end{equation}
where the Onsager coefficients satisfy $L_{11}\geq 0$, $L_{22}\geq 0$, $L_{11}L_{22}-L_{12}L_{21}\geq 0$ and $L_{12}=L_{21}$. Furthermore, the tight-coupling condition $L_{12}^{2}=L_{21}^{2}=L_{11}L_{22}$ leads to \begin{equation}J_2/J_1=L_{12}/L_{11}.\label{eq-tightc2}\end{equation}

If we take $J_2\equiv \dot{Q}_{c}$, the COP can be expressed as \begin{equation}\varepsilon\equiv\dot{Q}_{c}/{P}={L_{21}}/{TL_{11}X_{1}}\label{eq-chireg2}\end{equation} with the consideration of Eq.~(\ref{eq-tightc2}) and $P=TJ_1 X_1$. From Eq.~(\ref{eq-chireg2}) we solve $X_{1}={L_{12}}/TL_{11}{\varepsilon}$. Substituting it into the target function, we obtain
\begin{equation}\chi\equiv\varepsilon\dot{Q}_{c}=L_{22}(1+TX_{2}\varepsilon)/T.\end{equation}
It is easy to see that $\chi$ takes maximum when
\begin{equation}\varepsilon_\ast=0\label{eq-copMaxchiqc}\end{equation}
due to $X_2\equiv 1/T_h -1/T_c <0$.

Similarly, if we take $J_2\equiv \dot{Q}_{h}$, the COP can be expressed as
\begin{equation}\varepsilon\equiv\dot{Q}_{c}/{P}={L_{21}}/{TL_{11}X_{1}}-1\label{eq-chireg3}\end{equation}  with the consideration of $P=TJ_1 X_1$, $\dot{Q}_{c}=\dot{Q}_{h}-P$, and Eq.~(\ref{eq-tightc2}).
From Eq.~(\ref{eq-chireg3}) we have $X_{1}={L_{12}}/TL_{11}{(\varepsilon+1)}$.
Further, the target function is transformed into
\begin{equation}\chi\equiv\varepsilon\dot{Q}_{c}=L_{22}\varepsilon^{2}[1+TX_{2}(\varepsilon+1)]/T(\varepsilon+1)^2.\label{eq-chireg4}\end{equation}
Maximizing $\chi$ with respect to $\varepsilon$ (equivalently to $X_1$), we find that the COP at maximum $\chi$ satisfies $\varepsilon_\ast=\sqrt{{1}/{4}-2/TX_2}-{3}/{2}$. For small relative temperature difference $\varepsilon_C^{-1}\equiv(T_h-T_c)/T_c\ll 1$, we have $T\simeq T_c\simeq T_h$, and then Eq.~(\ref{eq-chireg4}) is transformed into \begin{equation}\varepsilon_\ast\simeq\sqrt{2\varepsilon_C}\label{eq-copMaxchiqh},\end{equation}
which is quite different from Eq.~(\ref{eq-copMaxchiqc}) by taking $J_2\equiv \dot{Q}_{c}$.

\section{Relationship between our picture and Prigogine's idea\label{sec-Prig}}
First, we will retype some text on pages 19-20 in Ref.~\cite{Prigoginebook}.

[begin]We consider again a system consisting of two closed phases, I and II, maintained respectively at uniform temperature $T^\mathrm{I}$ and $T^\mathrm{II}$. Applying formula (3.7) or (3.8) to
each phase, we have for the whole system, entropy
being an extensive variable,
$$\mathrm{d}S = \mathrm{d}S^\mathrm{I}+\mathrm{d}S^\mathrm{II} \eqno{(3.9)}$$

We now split the heat received by each phase into two parts (cf. 2.22)
$${\mathrm{d}}^\mathrm{I}Q={\mathrm{d}}_i^\mathrm{I}Q+{\mathrm{d}}_e^\mathrm{I}Q,~ {\mathrm{d}}^\mathrm{II}Q={\mathrm{d}}_i^\mathrm{II}Q+{\mathrm{d}}_e^\mathrm{II}Q \eqno{(3.10)}$$
where ${\mathrm{d}}_i^\mathrm{I}Q$ is the heat received by phase I from II, and $ {\mathrm{d}}_e^\mathrm{I}Q$ the heat supplied to phase I from the outside. Taking account of (2.24), we have for the entropy change of the whole system
$$\qquad\quad\mathrm{d}S =\frac{ \mathrm{d}^\mathrm{I}Q}{T^\mathrm{I}}+\frac{\mathrm{d}^\mathrm{II}Q}{T^\mathrm{II}} \qquad\qquad\qquad\qquad\eqno{(3.11)}$$
$$\qquad\qquad\quad=\frac{ \mathrm{d}_e^\mathrm{I}Q}{T^\mathrm{I}}+\frac{\mathrm{d}_e^\mathrm{II}Q}{T^\mathrm{II}}+\mathrm{d}_i^\mathrm{I}Q\left(\frac{1}{T^\mathrm{I}}-\frac{1}{T^\mathrm{II}}\right) \eqno{(3.12)}$$
In agreement with (3.3) the entropy change consists of
two parts.[end]

Let us consider an extended system including two heat baths and the engine. Let the cold bath and the hot bath, respectively, correspond to phase I and phase II in the book by Prigogine. There is no entropy change for the cyclic engine in each cycle or the autonomous engine in the steady state. Thus according to Prigogine's Eq.~(3.12) and the picture in Fig.~\ref{fig-engwh}(b), the entropy change per unit time of the extended system can be expressed in canonical form (\ref{eq-entcanon}) provided that we use the corresponding relations $\mathrm{d}_i^\mathrm{I}Q/\mathrm{d}t\leftrightarrow J_t $, $\mathrm{d}_e^\mathrm{I}Q/\mathrm{d}t\leftrightarrow -s_c\dot{W}$, and
$\mathrm{d}_e^\mathrm{II}Q/\mathrm{d}t\leftrightarrow -s_h\dot{W}$, where the minus signs represent the energy outflow from each phase since the physical quantities $\dot{Q}_c$, $\dot{Q}_h$ and $\dot{W}$ take their absolute values in the present work.

\section{Mapping a thermoelectric generator into our refined generic model\label{sec-thermEG}}

Apertet \emph{et al.} investigated an autonomous thermoelectric generator in recent work~\cite{ApertetPRE13}. Assume that $N$ electrons flow through the generator during time interval $\tau$. According to Eq.~(19) in their work~\cite{ApertetPRE13}, the heat absorbed from the hot bath or released into the cold bath per unit time can be expressed as
\begin{equation}\begin{array}{l}\dot{Q}_{h}={\alpha T_{h} Ne}/{\tau}-\gamma{RN^{2}e^{2}}/{\tau^{2}},\\
\dot{Q}_{c}={\alpha T_{c} Ne}/{\tau}+(1-\gamma){RN^{2}e^{2}}/{\tau^{2}},\end{array}\label{eq-thermoelec1}\end{equation}
respectively, where $\alpha$ is the Seebeck coefficient. The symbols $R$ and $e$ represent the electric resistance of the generator and the elementary electric charge, respectively. The power output may be expressed as
\begin{equation}\dot{W}\equiv\dot{Q}_{h}-\dot{Q}_{c}=\frac{\alpha Ne}{\tau}(T_{h}-T_{c})-\frac{RN^{2}e^{2}}{\tau^{2}},\label{eq-thermoelec2}\end{equation}

From Eq.~(\ref{eq-wtthflux}) the weighted thermal flux can be further expressed as $J_{t}=(s_{c}T_{h}+s_{h}T_{c})\alpha Ne/\tau+(s_{h}-\gamma)RN^{2}e^{2}/\tau^{2}$. To keep the leading term, we require $s_{h}-\gamma=0$. With the consideration of $s_{h}+s_{c}=1$, we obtain the weighted parameters:
\begin{equation}s_{h}=\gamma,~s_{c}=1-\gamma.\end{equation}
Then the weighted reciprocal of temperature and the weighted thermal flux can be expressed as
\begin{equation}\beta=\frac{\gamma T_{c}+(1-\gamma)T_{h}}{T_{c}T_{h}},\label{eq-thermoelec3}\end{equation}
and
\begin{equation}J_{t}=\frac{\alpha Ne\beta T_{c}T_{h}}{\tau},\label{eq-thermoelec4}\end{equation}
respectively. Obviously, the thermal flux $J_{t}$ couples tightly with the ``mechanical" flux $J_{m}\equiv 1/\tau$. With consideration of definitions (\ref{eq-wtthforce}) and (\ref{eq-mechfluxforce}), we derive
\begin{equation}J_{m}=\frac{T_cT_h}{[\gamma T_{c}+(1-\gamma)T_{h}]RN^{2}e^{2}}X_{m}+\frac{\alpha T_{c}T_{h}}{RNe}X_{t},\label{eq-thermoelec5}\end{equation}
from Eqs.~(\ref{eq-thermoelec2}) and (\ref{eq-thermoelec3}). Using Eqs.~(\ref{eq-linearcons}) and (\ref{eq-thermoelec3})--(\ref{eq-thermoelec5}), we obtain the Onsager coefficients
\begin{eqnarray}
&&L_{mm}=\frac{T_{c}T_{h}}{[\gamma T_{c}+(1-\gamma)T_{h}]RN^{2}e^{2}},\nonumber \\
&&L_{mt}=L_{tm}={\alpha T_{h}T_{c}}/{RNe},                                  \label{eq-Onsagerthermalelec}\\
&&L_{tt}=\frac{\alpha^{2}T_{c}T_{h}[\gamma T_{c}+(1-\gamma)T_{h}]}{R}.\nonumber
\end{eqnarray}

Thus thermoelectric generator mentioned above can be strictly mapped into our refined generic model. We should emphasize that the tight-coupling condition can be broken when the thermoelectric generator is put in a magnetic field~\cite{ApertetPRE13,Brandner2013,xzWang2013}. In this case, we need to revise our discussion in Sec.~\ref{sec-empheateng}.

\section{Expansion of mechanical flux up to the higher order for the Feynman ratchet\label{sec-FeynH}}
Let us continue to expand the net rate for the Feynman ratchet up to the higher order terms. First, the net rate (\ref{eq-fluxFeym}) can be transformed into
\begin{equation}r=r_{0}\mathrm{e}^{-\beta _{c}( \epsilon -z\theta _{c}) }[ \mathrm{e}^{(
\beta _{c}-\beta _{h}) \epsilon -z( \beta _{c}\theta _{c}+\beta
_{h}\theta _{h}) }-1],\end{equation}
which can be further transformed into
\begin{equation}r=r_{0}\mathrm{e}^{-\beta \epsilon }\mathrm{e}^{-[s_{h}\epsilon X_{t}+s_{c} X_{m}+\beta
^{-1} s_{h}s_{c}X_{t}X_{m}]}[\mathrm{e}^{(X_m+\epsilon X_t)}-1]\end{equation}
with the consideration of Eqs.~(\ref{eq-wttemp}), (\ref{eq-wtthforce}), (\ref{eq-wtnumFeym}) and (\ref{eq-JmFeym}).

With consideration of Eq.~(\ref{eq-JmFeym}), the mechanical force may be expressed as
\begin{eqnarray} J_{m}&=&r_{0}e^{-\beta\epsilon}[X_{m}+\epsilon X_{t}\nonumber\\ &+&(s_{h}/2-s_{c}/2)(X_{m}^{2}-\epsilon^{2}X_{t}^{2})]+O(X_{m}^{3}, X_{t}^{3}). ~~ \label{eq-FeynmanD3}\end{eqnarray}
For symmetric coupling, $s_h=s_c=1/2$, the quadratic order terms in above equation are vanishing.

Substituting Eq.~(\ref{eq-FeynmanD3}) into Eq.~(\ref{eq-maxp0}), we obtain the optimal mechanical force:
\begin{equation} X_{m}^{\ast}=-\frac{\epsilon}{2} X_{t}+\frac{ s_{h}-s_{c}}{16}\epsilon^{2}X_{t}^{2}+O(X_{t}^{3}) . \label{eq-FeynmanD4}\end{equation}
Substituting the above equation into Eq.~(\ref{eq-eta0}) with $\xi=\epsilon$, we obtain the efficiency at maximum power output
\begin{eqnarray} \eta^{\ast}&=&\frac{\eta_{C}}{2-s_{h}\eta_{C}+(s_{h}-s{c})\beta\epsilon\eta_{C}/4}+O(\eta_{C}^{3})\nonumber\\
&=&\frac{\eta_{C}}{2}+\left[\frac{1}{8}+\frac{(s_{h}-s_{c})(2- \beta\epsilon)}{16}\right]\eta_{C}^{2}+O(\eta_{C}^{3}).
 \label{eq-FeynmanD5}~~\end{eqnarray}
Obviously, when $s_{h}=s_{c}=1/2$ for the symmetric coupling, $\eta^{\ast}$ degenerates into the universal efficiency at maximum power output $(\eta_{C}/2+ \eta_{C}^{2}/8)$ up to the quadratic order.

\section{Thermodynamic fluxes and forces for a refrigerator\label{sec-appxe}}
A refrigerator can be described as a generic model which is similar to Fig.~\ref{fig-engwh} with all arrows being reversed. For a cyclic refrigerator in each cycle or an autonomous refrigerator in the steady state, the entropy production rate $\sigma$ may be expressed as:
\begin{equation}\sigma =\beta _{h}\dot{Q}_{h}- \beta _{c}
\dot{Q}_{c}.\label{eq-rfgEnP0}\end{equation}
With consideration of energy conservation (\ref{eq-Econst}), the above equation can be transformed into
\begin{equation}\sigma =\beta _{c}\dot{W}+\dot{Q}_{h}\left( \beta _{h}-\beta _{c}\right),\label{eq-rfgEnP1}\end{equation} or
\begin{equation}\sigma =\beta _{h}\dot{W}+\dot{Q}_{c}\left( \beta _{h}-\beta _{c}\right).\label{eq-regEnP2}\end{equation}

Introducing weighted parameters $s_c$ and $s_h$, the above two equations can be
transformed into
\begin{equation}\sigma =(s_{c}\beta _{c}+s_{h}\beta _{h}) \dot{W}+(s_{c}\dot{Q}_{h}+ s_{h}\dot{Q}_{c}) (\beta _{h}-\beta_{c}), \label{eq-rfgEnPf}\end{equation}
which enlightens us to define weighted reciprocal of temperature (\ref{eq-wttemp})
and weighted thermal flux (\ref{eq-wtthflux}).
The thermal force conjugated to $J_t$ may be defined as
\begin{equation}X_t\equiv \beta _{h}-\beta_{c}.\label{eq-rfgwtthforce}\end{equation}
The mechanical flux and force for the cyclic refrigerator can be expressed as
\begin{equation}J_m\equiv  1/t_0,\mathrm{~and~} X_m\equiv \beta W,\label{eq-rfgmechfluxforce}\end{equation}
respectively. For the autonomous refrigerator operating in the steady state, the mechanical flux and force may be expressed as
\begin{equation}J_m\equiv  r,\mathrm{~and~} X_m\equiv  \beta w,\label{eq-rfgmechfluxforce1}\end{equation}
respectively. Note that the signs of generalized thermodynamic forces $X_m$ and $X_t$ for the refrigerator are exactly opposite of those for the heat engine.

Finally, the entropy production rate, Eq.~(\ref{eq-rfgEnPf}), can be written in canonical form (\ref{eq-entcanon}) with consideration of Eqs.~(\ref{eq-wttemp}), (\ref{eq-wtthflux}), and (\ref{eq-rfgwtthforce})--(\ref{eq-rfgmechfluxforce1}).

\section{Curzon-Ahlborn endoreversible heat engine: quasi-tight-coupling engine\label{sec-appxf}}

The Curzon-Ahlborn heat engine~\cite{Curzon1975} undergoes a Carnot-like cycle consisting of two isothermal processes and two adiabatic processes.
In the isothermal expansion process, the working substance is in contact with a hot bath at temperature $T_{h}$ and absorbs heat $Q_{h}$ from that bath during time interval $t_{h}$. The effective temperature of the work substance is assumed to be $T_{he}$ ($T_{he}<T_{h}$). The variation of entropy in this process is assumed to be $\Delta S$.
In the isothermal compression process, the working substance is in contact with a cold bath at temperature $T_{c}$ and releases heat $Q_{c}$ to that bath during time interval $t_{c}$. The effective temperature of the work substance is assumed to be $T_{ce}$ ($T_{ce}>T_{c}$).
In two adiabatic processes, both the heat exchange and the variation of entropy are vanishing. Assuming the time for completing two adiabatic processes is negligible relative to $t_{c}$ and $t_{h}$. Thus the total time for completing the whole cycle can be written as $t_0= t_{h}+t_{c}$.

Because the working substance and the heat baths have different temperatures during the isothermal processes, the heat transfers may be expressed as
\begin{eqnarray}Q_{h}=\kappa _{h}t_{h}(T_{h}-T_{he}),~Q_{c}=\kappa _{c}t_{c}(T_{ce}-T_{c}),\label{eq-CA1}\end{eqnarray}
where $\kappa_{h}$ and $\kappa_{c}$ are the heat conductivities. In Ref.~\cite{wangtu2012}, Wang and one of the present authors have proved that the endoreversible assumption proposed by Curzon-Ahlborn~\cite{Curzon1975} may be expressed as
\begin{eqnarray}\Delta S=Q_{h}/T_{he}=Q_c/T_{ce}.\label{eq-CA2}\end{eqnarray}
Introducing two parameters $\gamma_{h}\equiv \kappa_{h}t_{h}/t_0$ and $\gamma_{c}\equiv \kappa_{c}t_{c}/t_0$, we can derive
\begin{eqnarray}
{Q}_{h}=\frac{T_{h}\Delta S}{1+\Delta S/\gamma_{h}t_0},~
{Q}_{c}=\frac{T_{c}\Delta S}{1-\Delta S/\gamma_{c}t_0}
\label{eq-CA3}\end{eqnarray}
from Eqs.(\ref{eq-CA1}) and (\ref{eq-CA2}).

With consideration of $\dot{Q}_{h}={Q}_{h}/t_0$, $\dot{Q}_{c}={Q}_{c}/t_0$, and $J_m\equiv 1/t_0$, we derive the weighted thermal flux $J_{t}\equiv s_{c}\dot{Q}_{h}+s_{h}\dot{Q}_{c}=(s_{c}T_{h}+s_{h}T_{c})\Delta SJ_m+(-s_{c}T_{h}/\gamma _{h}+s_{h}T_{c}/\gamma _{c})(\Delta SJ_m)^{2}+(s_{c}T_{h}/\gamma _{h}^{2}+s_{h}T_{c}/\gamma _{c}^{2})(\Delta SJ_m)^{3}+O(\Delta SJ_m)^{4}$. According to the quadratic order term of $J_t$ is zero as explained in Secs.~\ref{sec-physmean} and \ref{sec-lowdispat}, we require $s_{h}T_{c}/\gamma _{c}-s_{c}T_{h}/\gamma _{h}=0$. With consideration of $s_{h}+s_{c}=1$, we obtain the weighted parameters
\begin{equation}s_{h}=\frac{T_{h}\gamma_{c}}{T_{h}\gamma_{c}+T_{c}\gamma_{h}},~s_{c}=\frac{T_{c}\gamma_{h}}{T_{h}\gamma_{c}+T_{c}\gamma_{h}}\label{eq-CA4}.\end{equation}
Then the weighted reciprocal of temperature $\beta$ and the weighted thermal flux $J_{t}$ can be further expressed as
\begin{equation}\beta=\frac{\gamma_{h}+\gamma_{c}}{T_{h}\gamma_{c}+T_{c}\gamma_{h}}\label{eq-CA5}\end{equation}
and
\begin{equation}
J_{t}=\xi J_{m}+\xi_{3}J_{m}^{3}+O(J_{m}^4)\label{eq-CA6}\end{equation}
with $\xi=T_{c}T_{h}\beta\Delta S$ and $\xi_{3}=T_{c}T_{h}\beta\Delta S^{3}/\gamma_{c}\gamma_{h}$. If we neglect the higher order terms beyond the third order, the thermal flux $J_t$ is tight-coupled with the mechanical flux $J_m$. With consideration of the higher order terms, $J_t$ is merely quasi-tight-coupled with $J_m$, In this sense, the Curzon-Ahlborn heat engine is called a quasi-tight-coupling engine.

From Eqs.(\ref{eq-meanings}), (\ref{eq-power1}) and (\ref{eq-CA6}), we obtain the efficiency of the heat engine:
\begin{eqnarray}\eta &=&\frac{\dot{W}}{J_{t}+s_{h}\dot{W}}=\frac{-X_{m}J_{m}\beta^{-1}/J_{t}}{1-s_{h}X_{m}J_{m}\beta^{-1}/J_{t}}\nonumber\\
&=&\frac{-X_{m}\beta^{-1}/\xi}{[1+\xi_{3}J_{m}^{2}/\xi+O(J_{m}^{3})]-s_{h}X_{m}\beta^{-1}/\xi}\nonumber\\
&=&\frac{-X_{m}\beta^{-1}/\xi}{1-s_{h}X_{m}\beta^{-1}/\xi}+O({X_{m}\beta^{-1}/\xi})^{3}
\label{eq-CA7}.\end{eqnarray}

Considering Eqs.~(\ref{eq-Econst}), (\ref{eq-wtthforce}), (\ref{eq-mechfluxforce}) and (\ref{eq-CA3}) with $J_{m}\equiv 1/t_0$, we have
\begin{eqnarray}X_{m}\equiv &-&\beta W=-\beta(Q_{h}-Q_{c})\nonumber\\
=&-&(T_{h}-T_{c})\beta\Delta S+(T_{h}/\gamma_{h}+T_{c}/\gamma_{c})\beta\Delta S^{2}J_{m}\nonumber\\
&-&(T_{h}/\gamma_{h}^{2}-T_{c}/\gamma_{c}^{2})\beta\Delta S^{3}J_{m}^{2}+O(J_{m}^{3}),
\label{eq-CA8}\end{eqnarray}
From which we solve
\begin{eqnarray}J_{m}&=&\frac{\gamma_{c}\gamma_{h}}{(\gamma_{c}+\gamma_{h})\Delta S^{2}}\left[X_{m}+\xi X_{t}+\frac{s_{h}-s_{c}}{\Delta S}(X_{m}+\xi X_{t})^2\right]\nonumber\\
&+&O(X_{m}^{3},X_{t}^{3}),
\label{eq-CA9}\end{eqnarray}
with $\xi=T_cT_h\beta\Delta S$.
Substituting the above equation into Eq.~(\ref{eq-maxp0}) and considering $T_cT_h\beta^2=1+O(X_t)$, we can derive the optimal mechanical force
\begin{equation}
X_{m}^{\ast}=-\frac{\xi}{2} X_{t}+\frac{\xi(s_{h}-s_{c})}{8\beta}X_{t}^{2}+O(X_{t}^{3})
\label{eq-CA12}.\end{equation}
Finally, substituting the above equation into Eq.~(\ref{eq-CA7}), we can obtain the efficiency at maximum power output
\begin{eqnarray}
\eta^{\ast}&=&\frac{\eta_{C}}{2-\eta_{C}/2}+O(\eta_{C}^{3})\nonumber\\
&=&\frac{\eta_C}{2}+\frac{\eta_C^2}{8}+O(\eta_{C}^{3}),
\label{eq-CA13}\end{eqnarray}
with consideration of Eq.~(\ref{eq-wtthforce}) and $\eta_{C}\equiv 1-T_{c}/T_{h}$.
It is surprising that, up to the quadratic order, the efficiency at maximum power output for the Curzon-Ahlborn heat engine is independent of the symmetry of couplings between the heat engine and two baths.

\section{Mapping minimally nonlinear irreversible heat engine into our refined generic model \label{sec-appxg}}

First, we will briefly introduce the minimally nonlinear irreversible model of heat engine~\cite{Izumida2012}. The minimally nonlinear irreversible heat engine can be applied to both steady-state and cyclic heat engines. The thermal flux and thermal force are defined as $J_{2}\equiv \dot{Q}_{h}$ and $X_{2}\equiv 1/{T_{c}}-1/{T_{h}}$, where $T_{c}$ (or $T_{h}$) denotes the temperature of the cold bath (or hot bath). $Q_{h}$ is the heat absorbed from the hot bath by the working substance. The mechanical flux and mechanical force are defined as $J_{1}\equiv \dot{x}$ and $X_{1}\equiv -F/{T_{c}}$ for steady-state heat engines (or $J_{1}\equiv 1/{t_{0}}$ and $X_{1}\equiv -W/{T_{c}}$ for cyclic heat engines). The dot denotes derivative with respect to time, and $t_{0}$ is the time interval to complete the cycle. Then the relations between flux and force can be described by extended Onsager relations
\begin{equation} J_{1}=L_{11}X_{1}+L_{12}X_{2},~J_{2}=L_{21}X_{1}+L_{22}X_{2}-\gamma_{h}J_{1}^{2}, \label{eq-mni1}\end{equation}
where $L_{11}\geq 0,~L_{11}L_{22}-L_{12}L_{21}\geq 0$ and $L_{12}=L_{21}$ still satisfied. $\gamma_{h}$ is assumed to be a positive constant.

With consideration of tight-coupling condition Eq.~(\ref{eq-tightc0}), the second terms of Eqs.~(14) and (15) in Ref.~\cite{Izumida2012} are vanishing. Thus, the absolute value of heat absorbed from the hot bath and released to the cold bath can be expressed as
\begin{equation} \dot{Q}_{h}=\frac{L_{21}}{L_{11}}J_{1}-\gamma_{h}J_{1}^{2},~\dot{Q}_{c}=\frac{L_{21}T_{c}}{L_{11}T_{h}}J_{1}+\gamma_{c}J_{1}^{2},\label{eq-mni2}\end{equation}
respectively, where $\gamma_{c}\equiv \frac{T_{c}}{L_{11}}-\gamma_{h}>0$.

Now, we will construct the mapping from minimally nonlinear irreversible heat engine into our refined generic model. The weighted thermal flux~(\ref{eq-wtthflux}) can be further expressed as $J_{t}=\frac{L_{21}}{L_{11}}(s_{c}+s_{h}\frac{T_{c}}{T_{h}})J_{1}+(s_{h}\gamma_{c}-s_{c}\gamma_{h})J_{1}^{2}$. To keep the leading term, we require $s_{h}T_{c}/L_{11}-\gamma_{c}=0$. With consideration of $s_{c}+s_{h}=1$, the weighted parameters may be expressed as
\begin{equation} s_{h}=\frac{\gamma_{h}}{\gamma_{c}+\gamma_{h}},~s_{c}=\frac{\gamma_{c}}{\gamma_{c}+\gamma_{h}}.\label{eq-mni3}\end{equation}
With these parameters, the weighted reciprocal of temperature and the weighted thermal flux can be expressed as
\begin{equation} \beta=\frac{\gamma_{h}/T_{h}+\gamma_{c}/T_{c}}{\gamma_{c}+\gamma_{h}},\label{eq-mni4}\end{equation}
and
\begin{equation} J_{t}=\frac{L_{21}}{L_{1}}\frac{\gamma_{c}T_{h}+\gamma_{h}T_{c}}{T_{h}(\gamma_{c}+\gamma_{h})}J_{1}=\frac{L_{21}}{L_{11}}T_{c}\beta J_{1}.\label{eq-mni5}\end{equation}
Further more, from Eq.~(\ref{eq-mni4}) we can derive the relationship between $X_{m}$ in refined generic model and $X_{1}$ in minimally nonlinear irreversible heat engine
\begin{equation} X_{m}=\beta T_{c}X_{1}=\frac{\gamma_{c}T_{h}+\gamma_{h}T_{c}}{T_{h}(\gamma_{c}+\gamma_{h})}X_{1},\label{eq-mni6}\end{equation}
with consideration of Eq.~(\ref{eq-mechfluxforce}). Thus, when we adopt the mapping rules as $X_{m}=\beta T_{c}X_{1}$, $X_{t}=X_{2}$, $J_{m}=J_{1}$ and $J_{t}=L_{21}\beta T_{c}J_{1}/{L_{11}}$, the relationship shown in Eq.~(\ref{eq-mni1}) can be transformed into Eq.~(\ref{eq-linearcons}) with the Onsager coefficients satisfying
\begin{eqnarray}
&&L_{mm}=\frac{T_{h}(\gamma_{c}+\gamma_{h})}{\gamma_{c}T_{h}+\gamma_{h}T_{c}}L_{11},\nonumber \\
&&L_{mt}=L_{tm}=L_{12}=L_{21},                                  \label{eq-Onsagerthermalelec}\\
&&L_{tt}=\frac{\gamma_{c}T_{h}+\gamma_{h}T_{c}}{T_{h}(\gamma_{c}+\gamma_{h})}L_{22}.\nonumber
\end{eqnarray}
Thus, the tight-coupling minimally nonlinear irreversible heat engine can be strictly mapped into refined generic model of heat engine.

\end{document}